\documentclass[a4paper,useAMS,usenatbib]{mnras}

\usepackage{graphicx}
\usepackage{amsmath}

\title[The Galaxy End Sequence]{The Galaxy End Sequence}
\author[Stephen Eales]{Stephen Eales$^{1}$\thanks{E-mail:
sae@astro.cf.ac.uk}, Pieter de Vis$^{2}$, Matthew W.L. Smith$^1$,
Kiran Appah$^{1}$, 
\newauthor
Laure Ciesla$^3$, Chris Duffield$^{1}$,
and Simon Schofield$^{1}$\\
$^{1}$School of Physics and Astronomy, Cardiff University, The Parade, Cardiff CF24 3AA, UK\\
$^{2}$Sterrenkundig Observatorium,Universiteit Gent, Krijgslaan 281 S9, B-9000 Gent, Belgium\\ 
$^3$University of Crete, Department of Physics, 71003 Heraklion, Greece}
\begin{document}
\date{Accepted by MNRAS}
\date{}

\maketitle

\label{firstpage}

\begin{abstract}

A common assumption is that galaxies fall in two distinct
regions on a plot of 
specific star-formation rate (SSFR) versus galaxy stellar mass:
a star-forming Galaxy Main Sequence (GMS) and a separate
region of `passive' or `red and dead galaxies'.
Starting from a volume-limited sample of nearby galaxies
designed to contain most of the stellar mass in this volume,
and thus being a fair representation of the Universe at the end of $\simeq$12 billion 
years of galaxy evolution,
we investigate
the distribution of galaxies in this diagram today.
We show that galaxies
follow a strongly curved extended GMS with
a steep negative slope at high galaxy stellar masses. There
is a gradual change in the morphologies of the galaxies along this distribution,
but there is no clear break between early-type and late-type galaxies.
Examining the other evidence that there are two distinct populations, we argue that the `red
sequence' is the result of the colours of galaxies changing very little below a critical
value of the SSFR, rather than implying a distinct
population of galaxies, and that {\it Herschel} observations, which show at
least half of early-type galaxies contain a cool interstellar medium, 
also imply continuity between early-type and late-type galaxies.
This picture of a unitary population of galaxies requires more gradual
evolutionary processes than the rapid quenching processes needed to to explain
two distinct populations. We challenge theorists to reproduce the properties
of this `Galaxy End Sequence'.

\end{abstract}


\section{Introduction}

During the last decade, astronomers have developed
a simple phenomenological model of galaxy
evolution.
In this model, 
star-forming galaxies lie on the `galaxy main sequence' (henceforth
GMS), a distinct
region in a plot of star-formation rate versus galaxy stellar mass
(e.g. Noeske et al. 2007; Daddi et al. 2007; Elbaz et al. 2007; Rodighiero
et al. 2011; Whitaker et al. 2012; Lee et al. 2015).
Over cosmic time, the GMS gradually moves downwards in star-formation rate,
decreasing by a factor of $\simeq$20 from a redshift of 2 to the current epoch
(Daddi et al. 2007), with the cause of the evolution in the GMS being
the gradual decrease in the gas content of 
galaxies (Tacconi et al. 2010; Dunne et al.
2011; Genzel et al. 2015; Scoville et al. 2016).
An individual galaxy evolves along the GMS until some process quenches
the star-formation, and the galaxy then moves rapidly (in cosmic
terms) across the diagram to the region occupied by `red and dead' or
`passive' galaxies. 

Much of the physics behind this phenomenological model
is unknown.
The GMS is not as narrow as the stellar main sequence, with
a dispersion in the logarithm of star-formation rate of $\simeq$0.2 (Speagle et
al. 2014), and
it is still debated whether the GMS has any physical significance
(Gladders et al.
2013; Abramson et al. 2016). 
Peng et al. (2010) have shown many of the statistical
properties of star-forming and passive galaxies can be explained
if both the star-formation rate 
and the probability of quenching are proportional to the galaxy's stellar
mass, but the physics behind both 
proportionalities are unknown.
Although it is clear that the increased star-formation rates
in high-redshift galaxies are largely due to their increased gas
content, there is also evidence that the star-formation efficiency
is increasing with redshift (Rowlands et al. 2014; Santini
et al. 2014; Genzel et al. 2015; Scoville et al. 2016), and
so either the physics of star formation or the properties of
the interstellar gas (Papadopoulos and Geach 2012) must
be changing with redshift in some unknown way. 

Another big unknown is the role played by galaxy morphology. Investigations using
imaging with the {\it Hubble Space Telescope} have shown that at high redshift
the galaxies on the GMS are mostly late-type galaxies, galaxies dominated
by disks, while the galaxies that have been quenched are mostly early-type
galaxies, galaxies dominated by bulges (Pannella
et al. 2009; Wuyts et al. 2011; 
Lang
et al. 2014; Whitaker et
al. 2015; Schreiber et al. 2016). Therefore, in this paradigm, the process that quenches the
star formation also has to change the galaxy's morphology. Possibilies include,
but are not restricted to, galaxy merging (Toomre 1977) and the rapid motion
of star-forming clumps towards the centre of the galaxy
(Noguchi 1999; Bournaud et al. 2007; Genzel et al.
2011, 2014). This transformative process is of great importance
for the galaxy population as a whole, since the bolometric energy output from
the different morphological classes implies that while $\simeq$80\% of stars
were formed in disk-dominated galaxies, only $\simeq$50\% of stars in the
universe today are in these systems (Eales et al. 2015).

In this paper, we take a different approach to previous studies of the GMS.
These studies have 
been designed to study the process of galaxy evolution, and it has therefore
been important to
include large numbers of
star-forming galaxies since star formation is one of the
main drivers of this evolution.
In this paper 
we consider 
the galaxy population that the evolutionary processes have
produced in the Universe today. Since the result of
$\simeq$12 billion years of star formation are the stellar masses
of galaxies today, we start from a survey
that, rather than being designed to include star-forming galaxies,
was designed to include most of the stellar mass in the
Universe today. The {\it Herschel} Reference Survey\footnote{The {\it Herschel} Reference Survey
(P.I. Eales) was a key project carried out with the {\it Herschel Space Observatory}.
Most of the data for the survey, the {\it Herschel} data and the observational data
in other wavebands, can be obtained from the {\it Herschel}
Database in Marseille (hedam.lam.fr). The specific set of multi-wavelength
photometry used in this paper can be obtained from Pieter De Vis (pieter.devis@pg.canterbury.ac.nz).}
(HRS; Boselli et al.
2010) is a volume-limited sample of 323 galaxies
selected in the near-infrared K-band, which we
show in this paper is equivalent to a selection based
on galaxy stellar mass.
We use the extensive
photometry that exists for this sample, from the ultraviolet to the
far-infrared, to estimate the star-formation rates and stellar
masses of the HRS galaxies. Using these estimates, we investigate
the distribution of these galaxies in a plot of specific
star-formation rate versus galaxy stellar mass, and for the
first time
we investigate how galaxy morphology
varies along this distribution. We show that the properties of
this distribution set a number of awkward challenges for theorists
and observers attempting to produce a comprehensive model
of galaxy evolution.

We assume a Hubble constant of $\rm 67.3\ km\ s^{-1}\ Mpc^{-1}$ (Planck Collaboration 2014). 

\section{The Herschel Reference Survey}

The {\it Herschel} Reference Survey consists of 323 galaxies with distances
between 15 and 25 Mpc and with a near-infrared K-band limit
of $K<8.7$ for early-type galaxies (E, S0 and S0a) and $K<12$ for late-type
galaxies (Sa-Sd-Im-BCD). More details of the selection criteria are
given by Boselli et al. (2010). 
The sample was designed to be a volume-limited sample of galaxies
selected on the basis of stellar mass and suitable for an observing
programme with the {\it Herschel Space Observatory} (Pilbratt et al. 2010).
The flux limits were applied in the K-band to
(a) minimize the effect of dust and (b) produce a sample effectively
selected on the basis of galaxy stellar mass. 
The different flux limits for early-type
and late-type galaxies were chosen to avoid the sample being dominated
by low-mass early-type galaxies, which would have been hard to detect
with {\it Herschel}.
In Appendix A we investigate how much of the stellar
mass in the volume of space covered by the survey is actually
contained in the HRS galaxies.
One bias in the survey is that
the HRS volume contains the Virgo Cluster, and so the 
HRS is somewhat biased towards galaxies in denser environments.

\section{MAGPHYS}

The HRS now has high-quality photometry
in 21 photometric bands, including 
photometric measurements with GALEX (Cortese et al. 2012a),
SDSS (Cortese et al. 2012a), 2MASS (Skrutskie et al. 2006),
{\it Spitzer}/IRAC (Sheth et al. 2010), 
WISE (Ciesla et al. 2014),
Spitzer/MIPS (Bendo et al. 2012),
{\it Herschel}/PACS (Cortese et al. 2014) and {\it Herschel}/SPIRE (Ciesla et al.
2012). The quality of the photometry makes the HRS ideal for
the application of a galaxy modelling program such as
MAGPHYS (Da Cunha et al. 2008). Very briefly, MAGPHYS is based
on a simple model of the phases of the ISM. The program generates
50,000 possible models of the spectral energy distribution
of the unobscured stellar population, ultimately based on
the stellar synthesis models of Bruzual and Charlot (2003), and
the same number of models of the dust emission from the
interstellar medium. By linking the two sets of models
using a dust obscuration model that balances the radiation
absorbed at the shorter wavelengths with the energy emitted
in the infrared, the program generates a very large number
of templates which are fitted to the photometric measurements.
From the quality of the fits between the templates and the
measurements, the program produces probability distributions
for many of the important global properties of each
galaxy, such as the star-formation rate, stellar mass, total
mass of dust etc. The initial mass function on which
MAGPHYS is based is that of Chabrier (2003).

\begin{figure}
\includegraphics[width=64mm]{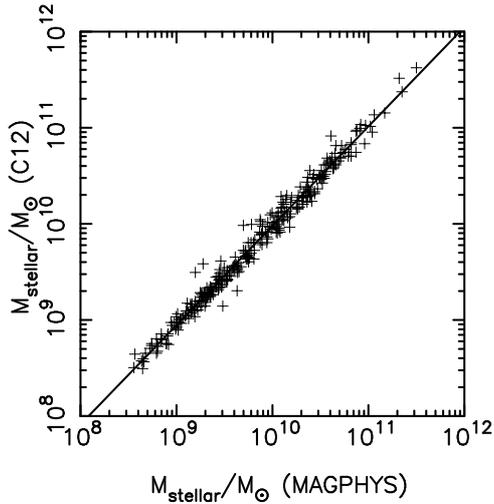}
  \caption{The stellar mass of an HRS galaxy estimated
by MAGPHYS plotted against the stellar mass estimated
by Cortese et al. (2012b). The latter estimates have been
corrected to the value of the Hubble Constant we use in this
paper. The line is the line that best fits the data
and is given by $log_{10}M_*(C12) = 1.034 log_{10}M_*(MAGPHYS)
-0.36$.
}
\end{figure}

De Vis et al. (2016) give a detailed description of
the application of MAGPHYS
to the HRS. 
De Vis et al. did not appy MAGPHYS to
four sources (HRS 138, 150, 183 and 241)
because the appearance of their spectral energy distributions indicated that
either the dust is heated by an AGN or a hot X-ray halo
or part of the far-infrared emssion
is from 
synchrotron radiation, neither of which
is included in the MAGPHYS model.
As our estimate of each galaxy
parameter, we have used the median value
from the probability distribution produced
by MAGPHYS for that parameter. Our estimate
of the star-formation rate in each galaxy corresponds to
the average star-formation rate over the last $10^8$ years.

As a basic check on the estimates provided by this complex
model, we compared the MAGPHYS estimates
of the galaxy stellar mass with those estimated
using a different method.
Cortese et al. (2012b) estimated galaxy stellar masses for the
HRS galaxies from the i-band luminosities and a relation
between mass-to-light ratio and g-i colour from
Zibetti et al. (2009).
Figure 1 shows a comparison between these estimates
and the MAGPHYS estimates, after correcting the 
estimates from Cortese et al. to the value of the
Hubble Constant that we assume here. 
The best-fit line to the points (see figure caption) shows
that at
a galaxy stellar mass of 
$\rm 10^{10}\ M_{\odot}$ estimated by MAGPHYS, the galaxy stellar mass estimated
by Cortese et al. is about 5\% lower.
But apart from this systematic
discrepancy (unimportant for the purposes of this paper),
the agreement between the two sets of galaxy stellar masses is
remarkably good.

We do not have independent measurements of the star-formation rate
with which to test the MAGPHYS estimates for the HRS galaxies.
However, Davies et al. (2016) did an extensive comparison of
the
different methods that have been used
for estimating star-formation rates.
They estimated
the star-formation rate for 
$\simeq$4000 star-forming galaxies using 12
different methods.
As their `gold standard', they adopted their own model
based on the radiative transfer model of
Popescu et al. (2011).
They used this model to recalibrate the
empirical methods, such as the relationship between
H$\alpha$ luminosity and star-formation rate (see below).
After recalibration, they found, not surprisingly, that
the slope of the relationship between star-formation rate and
galaxy stellar mass was very similar for the different methods.
The relationship from the MAGPHYS method, which was not corrected,
has a very similar slope (their Figure 10), which gives us
reassurance that the use of MAGPHYS 
will not lead to a bias in the relationship between specific star-formation
rate and galaxy stellar mass.

\section{Results}
We have used the MAGPHYS estimates to show the
positions of all the HRS galaxies in a plot of
specific star-formation rate against galaxy stellar mass (Fig. 2)\footnote{The
data used in this plot can be obtained from the authors.}.
We have used the morphological classification of
each galaxy (Boselli et al. 2010) to divide the
galaxies into eight classes: a) E and E/S0; b)
S0; c)
S0a and Sa; d) Sab and Sb; e) Sbc; f)
Sc and Scd; g) Sd and Sdm; h) I, I0, Sm and Im. 
We chose these groups to represent steps along
the 
traditional
Hubble sequence, and the points in Fig. 2 have been coloured
to show the galaxies in the different classes.

\begin{figure*}
\centering
\includegraphics[width=140mm]{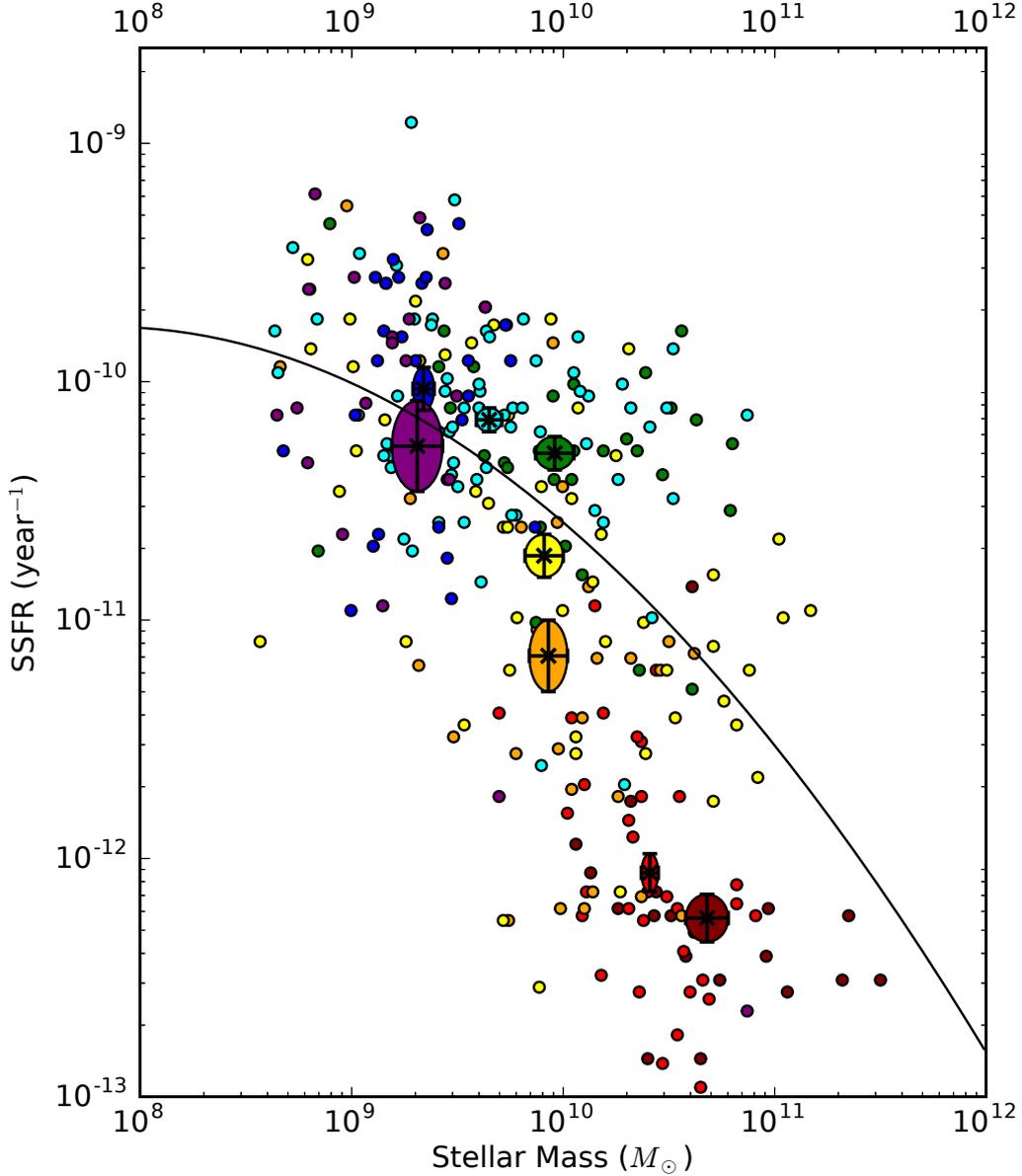}
  \caption{Plot of specific star-formation rate
(star-formation rate divided by stellar mass) plotted
against galaxy stellar mass.
We have used the morphological classification of
each galaxy (Boselli et al. 2010) to colour each galaxy
point using the following code: maroon - E and E/S0; red - S0;
orange - S0a and Sa; yellow - Sab and Sb; green - Sbc;
cyan - Sc and Scd; blue - Sd, Sdm; purple - I, I0, Sm and Im. 
The coloured ellipses show the $1\sigma$ error region on the mean
position for each morphological class (Table 1), with the colours
being the same as for the individual galaxies.
The line is the second-order polynomial that gives the
best fit to the data (see text for details) and
has the form:
$log_{10}SSFR=-10.59 - 0.76(log_{10}M_*-10.0)-0.176(log_{10}M_*-10.0)^2$.
}
\end{figure*}

Before considering the significance of this diagram,
there are two key observational issues to consider. The first
of these is the completeness of the diagram in terms of galaxy stellar mass.
In Appendix A we describe a detailed investigation
of the completeness of the HRS for both late-type and early-type
galaxies. We show that the HRS will include all
galaxies in the HRS volume down to galaxy stellar masses
of $\simeq 8\times10^8\ M_{\odot}$ for late-type galaxies
and $\simeq 2\times10^{10}\ M_{\odot}$ for early-type
galaxies.
Thus the HRS will have missed early-type galaxies in the
bottom left-hand corner of Fig. 2. However, there is very little
stellar mass in these omitted galaxies, and we show in Appendix A
that the mean galaxy
stellar mass of the early-type galaxies included in the HRS is very similar
to the mean galaxy stellar mass that would have been measured
if we had detected all early-type galaxies down to a galaxy stellar
masse of $10^8\ M_{\odot}$. Three of the four early-type galaxies
without MAGPHYS estimates (Section  3) have galaxy stellar masses
$>10^{11}\ M_{\odot}$, so their omission will have slightly reduced
the mean galaxy stellar mass, offsetting the first effect.

\begin{table}
\caption{Morphological Classes}
\begin{tabular}{cccc}
\hline
Morphological& No. & Mean $\rm log_{10}(M_*)$ & Mean $\rm log_{10}(SSFR)$ \\
class &  & & \\
\hline
E,ES0 & 18 & 10.68$\pm$0.10 & -12.25$\pm$0.10 \\
S0 & 34 & 10.41$\pm$0.04 & -12.06$\pm$0.08 \\
S0a,Sa & 32 & 9.93$\pm$0.09 & -11.15$\pm$0.15 \\
Sab,Sb & 55 & 9.91$\pm$0.09 & -10.73$\pm$0.09 \\
Sbc & 33 & 9.96$\pm$0.09 & -10.30$\pm$0.07 \\
Sc,Scd & 70 & 9.65$\pm$0.06 & -10.16$\pm$0.05 \\
Sd,Sdm & 29 & 9.34$\pm$0.05 & -10.03$\pm$0.09 \\
Sm,Im,I,I0 & 22 & 9.31$\pm$0.12 & -10.27$\pm$0.19 \\
\hline
\end{tabular}
\end{table}

The second issue is the accuracy of the specific star-formation rates (SSFR)
for galaxies with very low values of SSFR.
The median error given by MAGPHYS on $\rm log_{10}(SSFR)$ for galaxies with 
$\rm -12.0 < log_{10}(SSFR) < -11.0$ is
0.23 but increases to 0.39 for galaxies with 
$\rm log_{10}(SSFR) < -12.0$.
The values of SSFR for the galaxies in the lower
part of the distribution in Fig. 2 are thus very uncertain.
However, in this region of the diagram the distribution of galaxies is
almost vertical, and since the estimates of the
galaxy stellar mass are much more accurate than the SSFR estimates, it seems unlikely
that the large SSFR errors
are leading to 
an erroneous conclusion about the 
shape of the galaxy distribution.
Other methods for estimating the star-formation rate
also have great difficulty in providing useful measurements
for early-type galaxies with $\rm log_{10}(SSFR) < -12.0$
(e.g. Davis et al. 2014).

There are two things about the diagram that meet the eye.
The first is that the galaxy distribution appears curved. To test this statistically,
we fitted both a second-order polynomial and a straight line to the points
by minimimising the sum of chi-squared in the y-direction.
The curve in Fig. 2 is the best-fitting second-order polynomial.
The reduction in chi-squared obtained by fitting the polynomial
rather than the straight line
was $\Delta\chi^2 \simeq 56$. Since $\Delta\chi^2$ is distributed
as $\chi^2$ with one degree of freedom, the reduction in $\chi^2$ obtained
from fitting a second-order polynomial rather than a straight line is
highly significant, and thus there is strong statistical
evidence that the data is better represented by a curve than 
a straight line. We reached a similar conclusion from
minimising $\chi^2$ in the x-direction. 
The distribution in Fig. 2 is not the same as the star-forming main sequence
since we have plotted every galaxy in the figure rather than defining
a subset of star-forming galaxies. However, we note that there is other
evidence that the distribution of galaxies in this diagram is curved, whether
only star-forming galaxies are plotted (Whitaker
et al. 2014; Lee et al. 2015; Schreiber et al. 2016; Tomczak et al. 2016) 
or all galaxies
are plotted (Gavazzi et al. 2015).

Second, the galaxy morphologies appear to change systematically from the top left
to the bottom right of the distribution. To make this clearer, we have calculated the
mean values of $\rm log_{10}(SSFR)$ and $\rm log_{10}(stellar\ mass)$ for each morphological 
class. The
values are
given in Table 1 and plotted in Figure 2 with their $1\sigma$ error ellipses.
The diagram shows that as we move from top left to bottom right along the
galaxy distribution we also gradually move along the traditional Hubble sequence.

The HRS volume contains the Virgo Cluster. The HRS does not contain enough
galaxies to carry out an investigation of how the properties
of the galaxy population depend on environment (e.g. Casado et
al. 2015). However, we can
investigate whether the trends in Figure 2
are entirely due to the result of the galaxies in Virgo. To do this, we have removed
the HRS galaxies whose motions are claimed by Boselli et al. (2010) to be
dominated by peculiar motions in Virgo, which consists of the
galaxies in Virgo A and B,
the north and east clouds and the southern extension.
This reduces the HRS by $\simeq 50\%$ to 154 galaxies. In Figure 3 we
plot SSFR versus galaxy stellar mass for the remaining galaxies.
The same trends are visible.

\begin{figure}
\centering
\includegraphics[width=70mm]{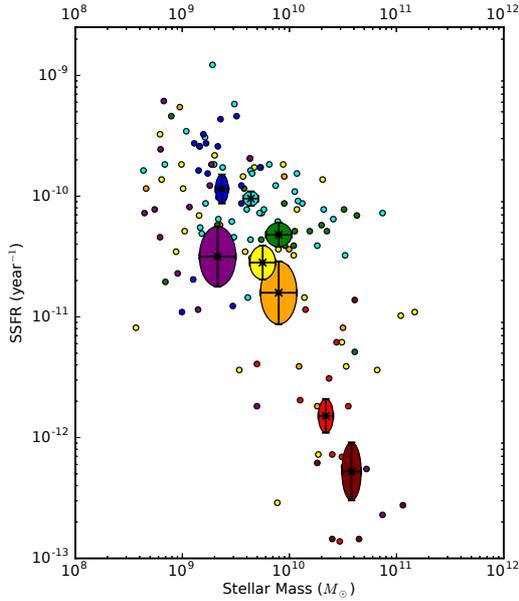}
  \caption{Plot of specific star-formation rate
(star-formation rate divided by galaxy stellar mass) plotted
against galaxy stellar mass for the
galaxies in the HRS outside the Virgo Cluster (see text
for details).
We have used the morphological classification of
each galaxy (Boselli et al. 2010) to colour each galaxy
point using the following code: maroon - E and E/S0; red - S0;
orange - S0a and Sa; yellow - Sab and Sb; green - Sbc;
cyan - Sc and Scd; blue - Sd, Sdm; purple - I, I0, Sm and Im.
The coloured ellipses show the $1\sigma$ error region on the mean
position for each morphological class with the colours
being the same as for the individual galaxies.
}
\end{figure}

We have compared our results with other attempts
to plot this diagram using samples from the
nearby Universe.
In our own investigation we have plotted all galaxies in this diagram 
rather than defining
a subset of star-forming galaxies, which makes it tricky
to compare our results with previous investigations of
the GMS.
Two techniques for defining the
star-forming GMS are to select galaxies with
an H$\alpha$ equivalent width above some critical value 
or to select galaxies with colours bluer than some critical value
(Speagle et al. 2014). The effect of both criteria is
to remove many of the
lower points in
Figure 2. For example, a critical H$\alpha$ equivalent width of
10\AA\ corresponds to an SSFR of $\rm 3 \times 10^{-11}\ year^{-1}$
(Casado et al. 2015), which would remove the
bottom half of Figure 2.
There is a similar effect with colour cuts. For example, Eales
et al. (2016) have plotted this diagram using $\simeq$3000 galaxies selected
from the {\it Herschel} ATLAS survey, showing that $\simeq$20-30\% of the galaxies
detected by Herschel
would have been classified as `passive galaxies' using optical criteria even though they
still have large reservoirs of gas and are still forming stars.

However, there are three recent studies with which we can compare our results.
First, Gavazzi et al. (2015) and Eales et al. (2016) have plotted this
diagram using samples selected from HI flux and submillimetre
flux, respectively, both making no attempt to remove `passive galaxies'. Their
results are shown in Fig. 4. In both cases, they found a distribution
with a similar to shape to
what we find here but offset to higher values
of SSFR. A possible explanation of the offset is if these two samples
are biased toward galaxies with large gas reservoirs and thus larger values
of SSFR. Fortunately, we can test this hypothesis because Eales et al. (2016)
investigated this effect by fitting a polynomial to their diagram
in two ways: not weighting the data points and weighting the datapoints by the
inverse of the accessible volume for each galaxy.
The two curved lines
in Figure 4 show the two fits. Whereas the line from the unweighted
fit is offset from our results, the line from the weighted fit follows
our distribution well. 

The third study is the H$\alpha$ study of Renzini and Peng
(2015). The authors attempted to overcome the problem of the rather arbitrary
criterion for
removing `passive galaxies' by fitting the
ridge line of the star-forming main sequence. We have plotted their results
over the range of galaxy stellar mass covered by their study, both the
relationship
they give for the relationship between SSFR and galaxy stellar mass and also
one we have recalculated from their relationship using the
the recalibration of the
relationship between H$\alpha$ luminosity
and star-formation rate derived by Davies et
al. (2016). 
The two relationships have a significant offset and
have different slopes, showing some of the uncertainty
of these studies. Chang et al. (2015)
have compared the star-formation rates used by
Renzini and Peng with estimates for the same galaxies from
MAGPHYS, finding that the latter are lower by
$\simeq0.22$ in $\rm log_{10}(SFR)$, 
which partly explains the offset between the
H$\alpha$ relationship and the HRS points.
Another possible explanation for an offset
is the rapid low-redshift evolution 
seen in the {\it Herschel} surveys (Dye et al. 2010; Dunne et
al. 2011; Eales et al. 2016; Marchetti et al.
2016). Marchetti et al. found that total far-infrared
luminosity, which is often used to estimate
the star-formation rate of a galaxy (Kennicutt 1998),
is proportional to $(1+z)^6$.
Even though the upper redshift limit of the sample
used by Renzini and Peng is only $0.085$, 
by this redshift, with such rapid evolution, the star-formation
rates of galaxies will have increased by
$0.21$ in $\rm log_{10}(SFR)$ relative to those
in a truly local sample such as the HRS.
The combination of the offset between the MAGPHYS and
H$\alpha$ estimates of the star-formation rate found by Chang et al. (2015) and the
effect of the rapid low-redshift evolution is probably enough
to bring the H$\alpha$ GMS into reasonable agreement with
the HRS GMS.

Given the uncertainties, there seems reasonable agreement between the different
studies of the GMS in the low-redshift Universe.

\begin{figure}
\centering
\includegraphics[width=70mm]{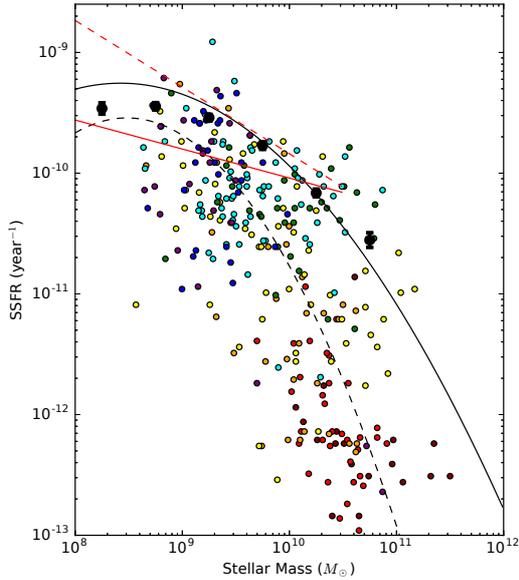}
  \caption{Plot of specific star-formation rate
(star-formation rate divided by galaxy stellar mass) plotted
against galaxy stellar mass for the
galaxies in the HRS. The key for the meaning of
the colours of the points is given
in the caption to Fig. 2.
The large black points show the results
for the HI-selected sample of Gavazzi et al. (2015).
The solid black line shows the fit to the $\simeq$3000
galaxies selected from the {\it Herschel} ATLAS survey (Eales
et al. 2016). The dashed line show the fit to the datapoints
from the same survey but with the contribution of
each galaxy to the fit weighted by the inverse of
the galaxy's accessible
volume. The red line shows the fit to the star-forming main sequence
from Renzini and Peng (2015). The dashed red line shows the fit for
the same sample after it has been recalculated
using the
recalibration of the relationship between H$\alpha$ luminosity
and star-formation
rate from Davies et al. (2016).
}
\end{figure}

\section{Discussion}

Our starting point in this investigation
was very different from investigations of the
star-forming GMS, because we are interested in what is left in the Universe
after $\simeq$12 billion years of star formation rather than in the
galaxies producing the stars. We showed in the last section that
when a number of obvious and not-so-obvious effects are taken into account,
there is fairly good agreement from the
different studies of the distribution of galaxies in the plot of galaxy stellar mass versus
SSFR in the Universe today.
The two clear trends with morphology that are seen in Figure 2,
that the specific star-formation
rate falls and the galaxy stellar mass rises as one moves along the Hubble sequence
from late-type to early-type galaxies, are not new (Devereux and Hameed 1997;
Kennicutt 1998; Dressler 2007).
The only new things we have done are to show this information
on the standard diagram used to interpret galaxy evolution, 
and to start from a survey that provides a fairly complete census
of the stellar mass in the local volume.
The main importance of this diagram is that any viable model of
galaxy evolution must end up with this distribution of galaxy properties.
The main practical limitation of our work
is that although we have accurate estimates of the stellar masses
of the the early-type galaxies, their star-formation rates are often
uncertain. Future observations, in particular with the JWST, 
will
be necessary to delineate the lower part of the distribution.

Galaxy evolution is now often considered
in terms of two distinct populations: star-forming and `passive galaxies',
which are usually assumed to correspond to late-type galaxies
(henceforth LTGs) and early-type galaxies (henceforth ETGs).
However, 
there is no clear evidence
for this in 
Figure 2. There is a possible clump of ETGs in Fig. 2 at $\rm SSFR < 10^{-12}\ year^{-1}$
but this is where the values of SSFR are extemely inaccurate, and otherwise there
is a gradual change in morphology along the distribution. In 
H$\alpha$ surveys there is often also often a clump of points around 
an SSFR of $\rm \simeq10^{-12}\ year^{-1}$ 
(e.g. Renzini and Peng 2015), but such values of SSFR correspond to
extremely low values of the H$\alpha$ equivalent width ($\sim 1 \AA$),
and often the clump appears
to be a set of upper limits rather than measurements (Renzini and Peng 2015).
Establishing whether there are two distinct populations is, of course,
of great importance
for understanding the physics of galaxy evolution. The existence of two
populations is usually explained by a rapid quenching process (e.g. Peng et al. 2010)
whereas the existence of a single population with gradually changing properties
would require more gentle evolutionary processes. 

When one looks in the literature there is a surprising amount of 
recent evidence for a unitary population of galaxies.
For example, Casado et al. (2015), using H$\alpha$ equivalent width
as a measure of SSFR and optical luminosity as a measure of galaxy stellar
mass, find that field galaxies follow a similar smooth curved
distribution to the one we see here (their Figure 8).
The ATLAS$^{3D}$ survey, a volume-limited
survey of 260 nearby ETGs similar in many respects to
the HRS, has revealed
that 86\% of ETGs have the velocity field 
expected for 
a rotating disk (Emsellem et al. 2011), and for 
92\% of these `fast rotators'
there is also photometric evidence for a stellar disk (Krajnovic
et al. 2013).
Cappellari et al. (2013) have used the ATLAS$^{3D}$ results
to propose that
there is a gradual change in galaxy properties from LTGs to ETGs
rather than a dichotomy at the ETG/LTG boundary, with the only
exception being 
the 14\% of the ETGs that are `slow-rotators', for
which there is generally (but not always) no photometric evidence
for a stellar disk (Krajnovic et al. 2013).
Cortese et al. (2016), using integral-field spectroscopy
from the SAMI survey, have also recently argued that, based on their kinematic properties,
LTGs and fast-rotator ETGs form a continuous class of objects.

Other evidence comes from surveys of the interstellar medium (ISM)
in ETGs. The ATLAS$^{3D}$ survey has detected molecular
gas in 22\% of ETGs (Young et al. 2011), and interferometric observations
of the ETGs generally show a rotating disk similar to
what is seen in LTGs (Davis et al. 2013). To date, the most sensitive way of detecting
the ISM in ETGs has been to use
the {\it Herschel Space Observatory} to observe the continuum dust emission.
Smith et al. (2012) detected dust emission from 50\% of the HRS ETGs,
and {\it Herschel} observations of a much larger sample of ETGs drawn from
the ATLAS$^{3D}$ survey have detected a similar percentage 
(Smith et al. in preparation). 
The big increase
in detection rate from ground-based CO observations to {\it Herschel} observations
suggests that the common assumption that ETGs do not contain
a cool interstellar medium is largely a function of instrumental sensitivity, and that
if we had more sensitive instruments we would find an even higher fraction
of ETGs with evidence for a cool ISM. 
These recent observations of the ISM in ETGs therefore suggest that, although
ETGs generally contain less cool gas than LTGs, there is a gradual change 
in the properties of the
cool ISM along the Hubble sequence.

There are two obvious counter-arguments.
The first 
is the claim that 
the apparent continuity between the ISM properties of
ETGs and LTGs is misleading because the gas in ETGs
has mostly been acquired as the result of
recent mergers.
The evidence for this 
is that $\simeq$24\% of ETGs have a difference of $>$30 degrees
between the
position angles of the 
stellar and gas rotational axes
(Davis et al. 2011). 
Although this is clearly evidence that some gas in ETGs has been acquired from mergers, 
in the vast majority of the galaxies in 
the sample of
Davis et al. (33 out of 38) the gas is 
misaligned by less than 90 degrees, such that the 
gas rotation vector is still broadly pointing in the same direction as 
that of the stars (23 of the galaxies have the position
angles of the two axes being within 10 degrees of parallel, three
have the position angles within 10 degrees of anti-parallel). This 
suggests that the effect of gas 
acquired from mergers is mostly to perturb a pre-existing ISM: if 
the galaxy contained no gas before the merger, the gas rotation vector 
would have an equal probability of being misaligned by less than and more than 90 degrees. 
Numerical simulations show that the torque between the acquired gas 
and the quadrupolar gravitational potential of the stars will eventually bring the 
two rotational axes into alignment, whether parallel or anti-parallel, but this process 
acts over 
long timescales, typically $\sim$2Gyr (van de Voort et al. 2015). 

The second counter-argument is that ETGs form a narrow `red sequence' in colour
plots (e.g. Bell et al. 2004), suggesting that ETGs form a distinct separate population to
LTGs. We show in Appendix B that 
below a specific star-formation rate of $\rm \simeq5 \times 10^{-12}\ year^{-1}$ 
the relationship between colour
and specific star-formation rate is so weak that, even if galaxies have a uniform
distribution of specific star-formation rate, they will still follow
a narrow
red sequence in colour plots.

Our results and the other recent results described above suggest that
late-type galaxies and most early-types (with the possible exception
of the 14\% of ETGs that are slow rotators) occupy a single continuum.
In this picture, a rapid quenching process in which gas is
ejected from a late-type galaxy, converting it into a `red-and-dead', `passive',
or `quiescent' 
galaxy, is not required, instead some more gradual evolutionary process, which
we will call `slow quenching'. 

There are other observational results both in favour of
and against this hypothesis.
Casado et al. (2015) 
combined H$\alpha$ observations,
sensitive to very recent periods of star formation, with optical colours,
sensitive to periods of star formation at earlier times, with the aim of
looking for sudden changes in the star-formation rate that might indicate
rapid quenching. They found no evidence for any such 
process in low-density environments, although they did find
evidence for a rapid quenching process in dense environments.
Using a similar technique, Schawinski et al. (2014) found evidence
for slow quenching in late-type galaxies but rapid quenching in early-type
galaxies.
Peng, Maiolino and Cochrane (2015) have used the metallicity
distributions of star-forming and quiescent galaxies to
argue that the evolution from one population to the other
must have occurred over $\simeq$4 billion years - evidence
for slow quenching.
Schreiber et al. (2016) concluded that at high redshift the decrease in the
slope of the star-forming GMS at high galaxy
stellar masses can be explained by slow quenching.
Although the same problems of measuring accurate values for galaxies
with low values of SSFR exist at high redshift, there is some evidence
for two distinct populations in the results of Elbaz et al. (2007; their Fig. 17) 
and of Magnelli et al. (2014; their Fig. 2), supporting rapid quenching.
One possible way to reconcile all these results would be if
most galaxies evolve by gradual processes (slow quenching) with the
14\% of ETGs that are slow rotators being produced by a
rapid quenching process.

In summary, we have shown that after 12 billion years of evolution the
galaxies in the local volume
follow a curved distribution in a plot of SSFR versus galaxy stellar
mass, with galaxy morphology changing gradually along this
distribution. Any viable theory of galaxy evolution must be
able to reproduce the properties of this distribution.
Our results and other recent results cast doubt on the
idea that there are two distinct populations of galaxies and suggest
that most galaxies occupy a single continuum.
These results suggest that catastrophic quenching
is not required to explain the properties of most of the
galaxy population and (to steal a geological term) a more
uniformitarian approach is required.

\section*{Acknowledgments}

We thank the referee, Dr. Corentin Schreiber, for comments that improved the
paper significantly.
SAE 
thanks the UK Science and Technology Facilities Council for
funding under consolidated grant ST/K000926/1.
MS and SAE are grateful for
funding from the European Union Seventh Framework Programme  ([FP7/2007-2013]
[FP7/2007-2011])  under  grant
agreement No. 607254.
\bibliographystyle{mnras}

\appendix

\section[]{How much of the stellar mass in the local volume is contained
in the galaxies in the Herschel Reference Survey?}

The {\it Herschel} Reference Survey consists of 323 galaxies with distances
between 15 and 25 Mpc and with a near-infrared K-band limit
of $K_{lim} <8.7$ for early-type galaxies (E, S0 and S0a) and $K_{lim}<12$ for late-type
galaxies (Sa-Sd-Im-BCD). We have used the following method to
estimate how the completeness of the HRS in this volume of space
depends on the galaxy stellar mass.

Figure A1 shows the absolute K-band magnitude plotted against
the galaxy stellar mass estimate from the MAGPHYS model for the
HRS galaxies. The relation is tight and there is no obvious
difference in the relationship
between early-type and late-type galaxies. The straight line
in the figure, which is the best fit to the data and was obtained
by minimizing the root-mean-squared-deviations in absolute magnitude,
has the following equation: 

\smallskip
\begin{align}
M_K = -2.26 \times log_{10}(M_*) + 0.07 
\end{align}
\smallskip

The completeness of the HRS, $C(M_*)$, in this volume of space is
given by:

\smallskip
\begin{align}
C(M_*) = {
\int^{D_{max}(M_*)}_{15Mpc} D^2 dD
\over
\int^{25Mpc}_{15Mpc} D^2 dD}
\end{align}
\smallskip

\noindent in which
$D_{max}(M_*)$ is the lower of 
25 Mpc and
$D_{lim}$, where $D_{lim}$ is given
by

\smallskip
\begin{align}
D_{lim} = 10^{{K_{lim} - M_K(M_*) - 25 \over 2.5}} 
\end{align}
\smallskip

Figure A2 shows how the completeness of the HRS depends on galaxy stellar mass
for late-type galaxies (blue line) and early-type galaxies (red line).
The figure shows that the HRS is complete for late-type galaxies
for galaxy stellar
masses above $\simeq 8 \times 10^8\ M_{\odot}$ 
and for early-type galaxies for galaxy stellar masses above $\simeq
2 \times 10^{10}\ M_{\odot}$. The implication of
the completeness limit for early-type galaxies
is that the HRS will have missed early-type galaxies in the lower
left-hand corner of Fig. 2.

Is this a big concern? 
One way to address this
question is to consider what fraction of the total
mass in the population of early-type galaxies is included
in the galaxies detected in the HRS.
In making this estimate, we make the assumption that
the
galaxy stellar mass functions
given by Baldry et al. (2012) for optically-red and
optically-blue galaxies are appropriate
for early-type and late-type galaxies, respectively.
Suppose $f(M_*)$ is the fraction of
the total galaxy stellar mass in galaxies with $M>M_*$
that is enclosed in the HRS volume and is actually included in the
HRS galaxies. This is given by

\smallskip
\begin{align}
f(M_*) = 
{\int^{\infty}_{M_*} M_* \phi(M_*) \int^{D_{max}(M_*)}_{15Mpc} D^2 dD dM_*
\over
\int^{\infty}_{M_{*}} M_* \phi(M_*) dM_* \int^{25Mpc}_{15Mpc} D^2 dD}
\end{align}
\smallskip
\noindent in which $\phi(M_*)$ is the galaxy stellar mass function
for either early-type or late-type galaxies.

The results for early-type and late-type galaxies
are shown by the dashed red and blue
lines in Fig. A2. The figure shows that $\simeq$90\%
of the total stellar mass in early-type galaxies with galaxy stellar
masses $>10^8\ M_{\odot}$ is contained in the galaxies
actually detected in the HRS.
The reason this percentage is so high is because the
galaxy stellar mass function for early-type galaxies has a maximum
at a galaxy stellar mass of $\simeq10^{10.5}\ M_{\odot}$,
and so even though the HRS completely misses early-type
galaxies in the lower left-hand corner of Fig. 2 the actual
stellar mass contained in these omitted galaxies
is very small.

Another question to ask is whether the horizontal position of the lower part
of the galaxy distribution in Fig. 2, the part traced by early-type galaxies,
has been biased by the omission of low-mass early-types. 
To estimate this we have calculated, first, the mean galaxy stellar mass of
the early-type galaxies detected in the HRS and then, using the formalism
above, the mean mass that we would have measured if we had detected
all early-type galaxies with galaxy stellar masses $>10^8\ M_{\odot}$.
The two vertical lines in Fig. A2 show the
two estimates.
The lines are very close, showing that the omission of low-mass galaxies has not
distorted our perception of the position of the galaxy distribution.  

\begin{figure}
\centering
\includegraphics[width=64mm]{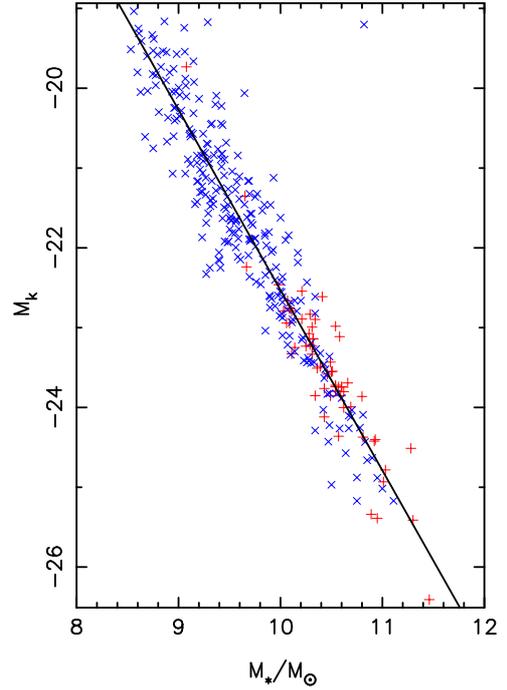}
  \caption{Plot of absolute K-band magnitude versus the galaxy stellar mass estimated
from the MAGPHYS model (\S 3) for the galaxies in the {\it Herschel} Reference Survey.
The blue crosses represent late-type galaxies and the red crosses early-type
galaxies. The equation of the line, which was fit to the points by
minimizing the root mean squared deviations in absolute magnitude, is 
$M_K = -2.26 \times log_{10}(M_*) + 0.07$. 
}
\end{figure}

\begin{figure}
\centering
\includegraphics[width=64mm]{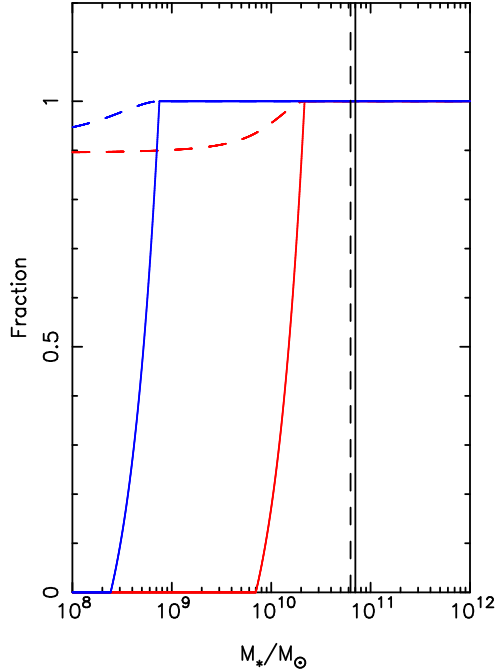}
  \caption{The fraction of galaxies in the HRS volume that are actually
included in the HRS plotted against galaxy stellar mass for late-type (solid blue
line) and early-type galaxies (solid red line).
The dashed lines show the predictions for the fraction of the
cumulative galaxy stellar
mass (the total galaxy stellar mass above the plotted
value) in the HRS volume that is included in the HRS late-type galaxies
(dashed blue line)
and early-type galaxies (dashed red line).
The vertical solid line shows the mean galaxy stellar mass of the early-type
galaxies in the HRS; the vertical dashed line shows
the predicted mean galaxy stellar mass if the HRS had detected all
early-type galaxies with galaxy stellar masses $>10^8\ M_{\odot}$.
}
\end{figure}

\section[]{The Cause of the Red Sequence}

In this section we show that a population of galaxies that is
uniformly distributed in the logarithm of specific star-formation (SSFR)
rate will naturally produce a `red sequence' in a colour diagram.
We started with a single stellar population from Bruzual and Charlot
(2003) with a Salpeter initial mass function and solar metallicity.
We then constructed a large number of models for the star-formation
history of a galaxy which were designed to produce the full range 
of SSFR observed in the universe today. 
These models have no great physical significance and the model from Bruzual
and Charlot (1983) was selected fairly randomly, since the goal
of the modelling was to see whether a red sequence could be produced even
for a uniform distribution in SSFR, not
that every possible galaxy-evolution model would produce a red sequence.

Our models of the star-formation history of a galaxy consisted of
two sets. In the first set, we assumed that the galaxy formed
12 Gyr ago and that the star-formation rate is proportional
to $e^{- {t \over \tau}}$, where $t$ is the time since the
formation of the galaxy and $\tau$ is a constant. In this
set, we used 1000 values of $\tau$ equally spread logarithmically
between 0.5 and 40 Gyr.
Since this set could not produce very high values
of SSFR, we also constructed a second set of models. In a model
in this second
set the star-formation history has two components. The first
component is an old component, with the exponential form
given above and a value of $\tau$ of 40 Gyr.
The second component is a burst of star formation that occurs
during the last $1.0\times10^8$ years. We constructed
a sequence of models, in which the star-formation rate in the
burst ranged from 0.01 to $100$ times the star-formation rate
at that time in the old component.

\begin{figure}
\centering
\includegraphics[width=64mm]{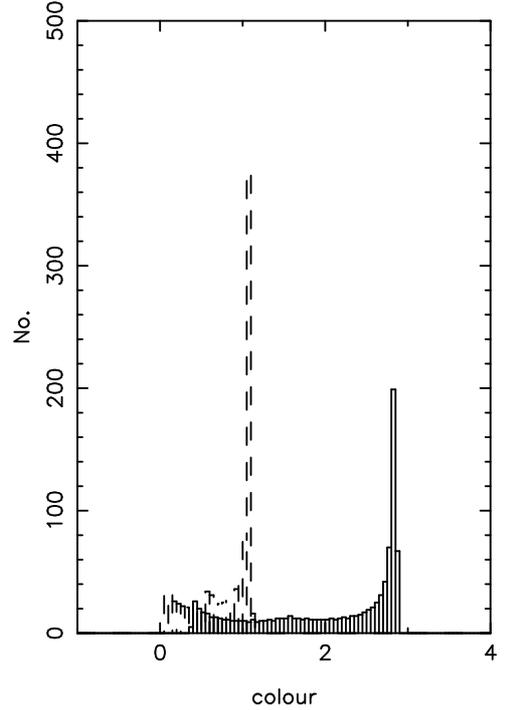}
  \caption{Distribution of colour predicted
by the models described in the text for a
population of galaxies
with $log_{10} SSFR$ uniformly
spread between -14.0 and -8.5. The solid line shows
the predicted distribution for $u-r$ colours
and the dashed line for $g-r$ colours.
}
\end{figure}

For each model, we calculated the colour and the SSFR at
the current epoch. From the two sets of models, we derived
a look-up table linking SSFR and colour. We then assumed
a population of galaxies with $\rm log_{10} SSFR$ uniformly
spread between -14.0 and -8.5. We then used the
look-up table to predict the distribution of colours
for this population. Fig. B1 show the predicted distribution
of $u-r$ and $g-r$ colour. In both cases, there
is a clear red sequence. The explanation
of the red sequence is straightforward. Below a SSFR of
$\rm \simeq 5 \times 10^{-12}\ year^{-1}$, colour depends very
weakly on SSFR, so all galaxies below
this value of SSFR have virtually the same colour. Therefore,
the existence of a red sequence is not a sign of a
single homogeneous class of galaxies but merely
that there are a large number of galaxies with values
of SSFR below this value.

It is worth pointing out that the corresponding colour limit
for a young stellar population could have produced a `blue sequence' but
in practice does not. The youngest possible population of stars would be
one containing only O stars. Such a population would have 
colours of $u-r \simeq -1.22$ and $g - r \simeq -0.58$. These
colours, however, are much bluer than the bluest galaxies in the
`blue cloud' (Baldry et al. 2012).

Another way that has been used to show the existence of
two distinct populations of galaxies - star-forming
and `passive' - is to measure the size
of the 4000\ \AA\ break from spectra from the Sloan Digital
Sky Survey (Kauffmann et al. 2003).
Kauffmann et al.
used the parameter $D_n$, which is the ratio of the
average flux density, $F_{\nu}$, in the wavelength
range $4000 < \lambda < 4100\ \AA$ to the average
flux density in the wavelength range $3850 < \lambda <
3950\ \AA$.
Galaxies fall nicely into two distinct areas
on a plot of $D_n$ verses stellar mass
(see Fig. 1 of Kauffmann et al.).
We repeated our modelling in exactly the same way as
before except that this time we calculated the relationship
between SSFR and $D_n$. Figure B2 shows the
predicted distribution of $D_n$ for a 
population of galaxies 
with $\rm log_{10} SSFR$ uniformly
spread between -14.0 and -8.5. There is a clear peak, again
a consequence of the weak dependence between $D_n$ and SSFR below
a value of SSFR of $\rm \simeq 2 \times 10^{-12}\ year^{-1}$.

\begin{figure}
\centering
\includegraphics[width=64mm]{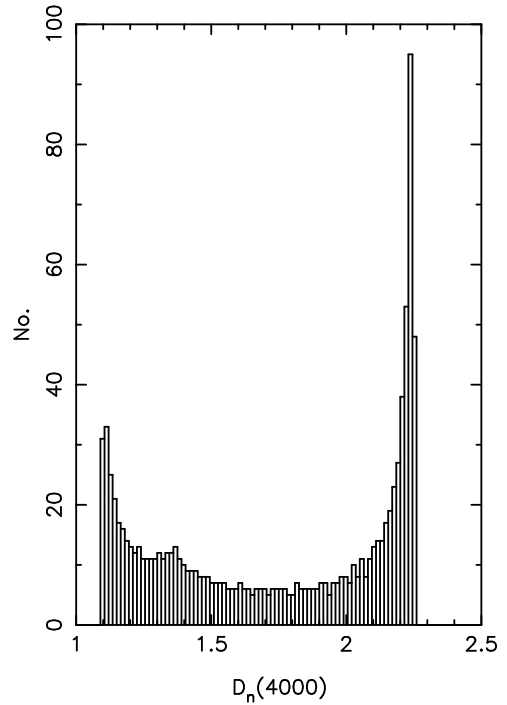}
  \caption{Distribution of the strength
of the 4000\ \AA\ break predicted
by the models described in the text for a
population of galaxies
with $log_{10} SSFR$ uniformly
spread between -14.0 and -8.5. 
}
\end{figure}

\label{lastpage}

\end{document}